\documentclass[aps,prb,showpacs,twocolumn,amsmath,groupedaddress,letterpaper]{revtex4} 
 
\usepackage{graphicx}% Include figure files 
\usepackage{bm}% bold math 
 
\def\beq{\begin{equation}} 
\def\eeq{\end{equation}} 
\def\beqn{\begin{eqnarray}} 
\def\eeqn{\end{eqnarray}} 
 
\begin{document} 
 
\title{Spin-dependent scattering in a silicon transistor} 
 
\author{Rog\'{e}rio de Sousa} \affiliation{Department of Physics and 
  Astronomy, University of Victoria, Victoria, BC V8W 3P6, Canada} 
\author{Cheuk Chi Lo} \author{Jeffrey Bokor} 
\affiliation{Department of Electrical Engineering and Computer 
  Sciences, University of California, Berkeley, CA 94720, USA} 
 
\date{\today} 
 
\begin{abstract}
  The scattering of conduction electrons off neutral donors depends
  sensitively on the relative orientation of their spin states. We
  present a theory of spin-dependent scattering in the two dimensional
  electron gas (2DEG) of field effect transistors.  Our theory shows
  that the scattering mechanism is dominated by virtual transitions to
  negatively ionized donor levels.  This effect translates into a
  source-drain current that always gets \emph{reduced} when donor
  spins are at resonance with a strong microwave field.  We propose a
  model for donor impurities interacting with conduction electrons in
  a silicon transistor,  and compare our explicit numerical calculations
  to electrically detected magnetic resonance
  (EDMR) experiments.  Remarkably, we show that EDMR is optimal for
  donors placed into a sweet spot located at a narrow depth window
  quite far from the 2DEG interface.  This allows significant
  optimization of spin signal intensity for the minimal number of
  donors placed into the sweet spot, enabling the development of
  single spin readout devices.  Our theory reveals an interesting
  dependence on conduction electron spin polarization $p_c$. As $p_c$
  increases upon spin injection, the EDMR amplitude first increases as
  $p_{c}^{2}$, and then saturates when a polarization threshold $p_T$
  is reached. These results show that it is possible to use EDMR as an
  in-situ probe of carrier spin polarization in silicon and
  other materials with weak spin-orbit coupling.
\end{abstract} 
\maketitle 
 
\section{Introduction}

Electron spins carry much promise as state variables for scaled
classical logic\cite{zutic04} and as quantum bits (qubits) in quantum
computer architectures.\cite{kane98} A key challenge of current
research in spin-based electronics (``spintronics'') and quantum
computation is to devise methods to probe spin polarization in
semiconductors with weak spin-orbit coupling,\cite{zutic06} like
silicon and silicon-germanium alloys.  Silicon combines many special
features that make its electronic spin a promising basis for classical
and quantum logic devices: Band structure properties such as weak
spin-orbit coupling and indirect band gap, and the possibility of
preparing a nuclear spin free environment lead to extremely long spin
relaxation times in comparison to other
semiconductors.\cite{tyryshkin03}
 
However, the same features that lead to long spin relaxation times 
also make spin detection more challenging.  Optical methods for spin 
detection such as Faraday and Kerr rotation have been used successfully 
in III-V semiconductors,\cite{kikkawa99,crooker05} but unfortunately, these 
methods are inapplicable to silicon and related materials, whose 
spin-selective optical transitions are extremely weak and ineffective. 
 
% Use 5inches for single column, use 3 inches for double. 
\begin{figure} 
\includegraphics[width=3in]{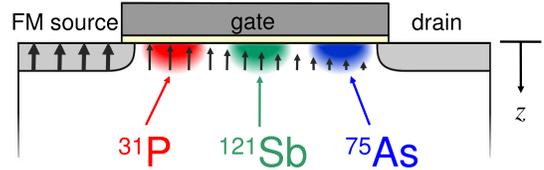} 
\caption{(Color online) Silicon transistor with a ferromagnetic
  source, inspired by the devices in [\onlinecite{jonker07,appelbaum07}]. Our
  theory shows that doping with different types of group V donor
  species along the transistor channel enables the determination of
  the conduction electron spin polarization distribution along the
  channel. Each donor species has a distinct set of hyperfine-split
  electron spin resonance frequencies, whose amplitude is directly
  proportional to the local carrier spin and donor electron spin
  polarizations.\label{device}}
\end{figure}

Recently, injection and detection of conduction electron spin 
polarization was demonstrated in silicon.  In Ref.~[\onlinecite{jonker07}], a Si 
device with a ferromagnetic (FM) source was 
 interfaced with GaAs to enable optical spin detection, 
while in [\onlinecite{appelbaum07}] a spin valve based on a 
 FM-Si interface was used. 
 
Both experiments achieve spin detection at the interface, but no local 
or ``in-situ'' spin detection. The interface limits spatial resolution 
and introduces additional scattering effects.  In studies with 
accumulation channel silicon transistors, spin-dependent 
scattering was demonstrated by electrical detection of magnetic 
resonance (EDMR).\cite{ghosh92,lo07}
 
In this article we present a theory of spin-dependent transport in the 
scattering of spin polarized conduction electrons in the two 
dimensional electron gas of field effect transistors.  Based on the 
scattering theory, we propose a donor-based approach to the problem of 
\emph{in-situ carrier spin detection in silicon}.  At the limit of a 
single donor impurity, this mechanism also provides a path for the 
readout of single donor spin states that does not require proximity to 
(or presence of) charge traps.\cite{kane98,xiao04,elzerman04,kaplan78}
 
Consider a silicon transistor with a series of different group V donor
impurities implanted along its conduction channel, Fig.~\ref{device}.
Spatially resolved characterization of carrier spin polarization is
obtained by measuring the source-drain current when each donor species
is in resonance with a microwave field. 
 
This is possible due to spin-dependent scattering (SDS), which relies
solely on the symmetry of impurity scattering
events.\cite{honig66,schmidt66,ghosh92} While this effect was believed to be
observed through EDMR in [\onlinecite{ghosh92}] amidst competing
heating effects, it was only recently that SDS was confirmed
experimentally using a small ensemble ($6 \times 10^6$) of implanted
antimony (Sb) donors.\cite{lo07} In the following, we present a general
microscopic theory of SDS, its efficiency and tunability in a
two dimensional electron gas (2DEG).  We show that the core physical
mechanisms underlying SDS are virtual transitions mediated by
negatively ionized donor levels.  Our results apply to all
semiconductors, but are particularly useful as a probe of spin
polarization in group IV and related materials with weak spin-orbit
coupling, whose spin detection is intrinsically challenging.  Our
physical optimization of this effect has exciting consequences for
spintronics and single donor electron or nuclear spin readout.\cite{sarovar07}

We remark that SDS is quite different from other EDMR experiments
based on spin-dependent recombination of electron-hole 
pairs.\cite{lepine72,kaplan78,christmann95,brandt04,stegner06,mccamey06,morley08}
As showed by Kaplan, Solomon, and Mott,\cite{kaplan78} spin-dependent
recombination has important contributions that \emph{do not} depend on
carrier spin polarization $p_c$. Hence it is difficult to use
recombination based EDMR as a probe of carrier spin polarization
$p_c$.

This article is organized as follows. Section~\ref{sdsedmr} presents a
general theory of SDS and its detection by EDMR in the case of thermal
equilibrium in an external magnetic field (no spin injection).
Section~\ref{numerics} describes our explicit numerical calculations,
together with a comparison between theory and two
experiments,\cite{ghosh92,lo07} and a discussion on how SDS can be
optimized with respect to donor impurity location.
Section~\ref{kondosection} discusses the validity of our perturbation
theory approach, by showing explicit calculations of the Kondo
temperature as a function of donor impurity location.
Section~\ref{spininjec} generalizes our theory to the case of
non-equilibrium spin injection, and discusses how EDMR scales with
carrier spin polarization. Section~\ref{conclusions} presents our
conclusions.

\begin{figure} 
\includegraphics[width=3in]{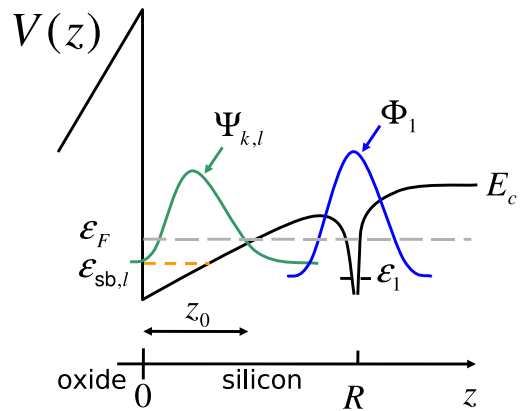} 
\caption{(Color online) Band diagram along the $z$ direction
  perpendicular to the 2DEG plane, as shown in Fig.~\ref{device}.
  Here $E_c$ is the conduction band edge, $\epsilon_F$ is the 2DEG
  Fermi energy, $\epsilon_{\rm{sb},l}$ is the ground state energy for
  the $l$-th subband, and $\epsilon_1$ is the ground state energy of a
  neutral donor (singly occupied).  We also show the donor impurity
  ground state wavefunction $\Phi_1$, and the conduction electron
  wavefunction $\Psi_{k,l}$ with its characteristic length $z_0$.  As
  we show below, only donors located at $R\gg z_0$ can be detected by
  EDMR.\label{band_diagram}}
\end{figure} 
 
\section{Theory of spin-dependent impurity scattering and electrically
  detected magnetic resonance\label{sdsedmr}}

Impurity scattering depends on whether the two-particle 
states formed by a conduction and an impurity electron are in a 
singlet (S) or triplet (T) configuration, 
\begin{equation} 
\Psi^{S/T}_{kl}=\frac{1}{\sqrt{2}} 
\left[\Psi_{kl}(\bm{r}_{1})\Phi_1(\bm{r}_{2}) 
\pm \Psi_{kl}(\bm{r}_{2})\Phi_1(\bm{r}_{1})\right]. 
\label{st} 
\end{equation} 
Here $\Psi_{kl}$ and $\Phi_1$ are orbital wavefunctions, respectively
of a conduction electron with momentum $k$ in the $l$-th subband, and
a localized donor impurity electron in the ground state.
Fig.~\ref{band_diagram} shows the band diagram and potential profile
for conduction and impurity wavefunctions in the 2DEG.

\begin{figure} 
\includegraphics[width=3in]{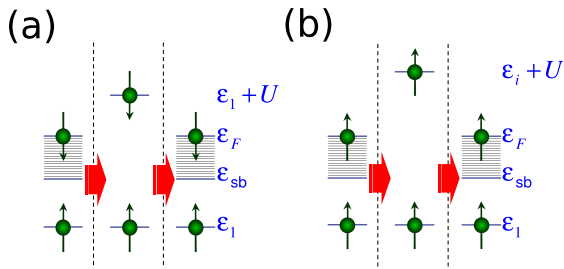} 
\caption{(Color online) Contributions to neutral impurity scattering. 
  (a) The largest individual contribution to singlet scattering is a 
  virtual transition to a negatively ionized donor level, where two 
  electrons occupy the same (ground) orbital donor level $\epsilon_1$. 
  Such a transition is symmetry forbidden for triplets due to Pauli 
  exclusion. (b) Triplet scattering occurs due to virtual transitions 
  into negatively ionized donor levels where one electron occupies the 
  level $\epsilon_1$, and the other electron occupies an excited 
  orbital state $\epsilon_i>\epsilon_1$. These same transitions into 
  excited orbital states are also allowed for singlets.\label{diagram}} 
\end{figure}

Access to additional channels for virtual scattering in the singlet
state (Fig.~\ref{diagram}) translates into distinct neutral impurity
scattering times for triplets vs. singlets, $\tau_{T}\neq \tau_{S}$.
With spin polarizations
$p=(p_{\uparrow}-p_{\downarrow})/(p_{\uparrow}+p_{\downarrow})$, for
conduction ($p_c$) and impurity ($p_i$) electrons, the occupation
fraction for singlets is given by $p_S=(1-p_i p_c)/4$, and the
fraction for triplets $p_T=1-p_S=(3 + p_i p_c)/4$.  Due to this
difference in singlet vs. triplet scattering times, the device current
is directly related to the spin polarizations according to
\begin{equation} 
I \propto \langle \tau\rangle = \left(p_S \tau_{S}+p_T \tau_{T}\right) 
\propto I_0 \left(1+\alpha  p_c p_i\right). 
\label{mu} 
\end{equation} 
Here we introduced the parameter $\alpha$ as a figure of merit for the
SDS effect.  Eq.~(\ref{mu}) forms the basis of the SDS mechanism of
EDMR detection, which only occurs when the carrier spin polarization
$p_c$ is non-zero. At thermal equilibrium, and in the low temperature ``degenerate limit''
($k_BT\ll \rm{Min}_l\{\epsilon_F-\epsilon_{\rm{sb},l}\}$), $p_c$ 
arises due to Pauli paramagnetism,
\begin{equation}
  p_c=\frac{n_\uparrow-n_\downarrow}{n_\uparrow+n_\downarrow}= 
  \frac{g\mu B}{2\left[\epsilon_F-\frac{1}{2}(\epsilon_{\rm{sb,1}}
          +\epsilon_{\rm{sb,2}})\right]}.
\end{equation}
Here $n_{\sigma}$, $\sigma=\uparrow,\downarrow$ is the electron
density for spin species $\sigma$, $B$ is the external magnetic field,
$g\mu/\hbar=\gamma_e=1.8\times 10^{7}$~(sG)$^{-1}$ is the free
electron spin gyromagnetic ratio, and
$\epsilon_{\rm{sb},1}$, $\epsilon_{\rm{sb},2}$ are the 2DEG subband
ground state energies.

Under microwave irradiation, the donor spin transition rates are given
by $W\equiv W_{\uparrow}=W_\downarrow$ (the subscript refers to the
initial state of the transition),
\begin{equation}
  W(\omega)= 2\pi \left|\gamma_e B_{\perp}\right|^2 
\frac{1}{\pi T_{2}^{*}}\frac{1}{(\omega-\omega_m)^{2}+\left(\frac{1}{T_{2}^{*}}\right)^{2}},
\label{weqn}
\end{equation}
where $B_{\perp}$ is the amplitude of the microwave
field, $\omega_m= \gamma_e (B+h_m)$ is the resonance frequency of
the donor with nuclear spin state $m=-I,-I+1,\ldots, +I$ ($h_m$ is a
hyperfine shift), and $\frac{1}{T_{2}^{*}}$ is the inhomogeneous linewidth. 
The microwave rate $W$ competes with the donor spin-flip rate
$\frac{1}{T_1}=\Gamma_{\uparrow}+\Gamma_{\downarrow}$ in order to determine a 
steady state impurity spin polarization,
\begin{equation}
p_i=\frac{p_{i0}}{1+2T_1W},
\label{pi}
\end{equation}
where $p_{i0}=\tanh{\left(\frac{g\mu B}{2k_BT}\right)}$ is
the equilibrium donor spin polarization. 

When donor atoms are at resonance ($\omega=\omega_m$), and satisfy the
saturation condition $(\gamma_e B_{\perp})^{2}\gtrsim 1/(T_1
T^{*}_{2})$, their spin polarization vanishes, $p_i \approx 0$,
effectively \emph{reducing the device current by $\Delta I/I_0 \approx
  \alpha p_{c}p_{i0}$}. In the ``clean limit'', when no other
scattering sources are competing with donor impurity scattering, the
figure of merit $\alpha$ is given by
\begin{equation}
\alpha=\frac{\sum_l \left(\tau_{T,l}-\tau_{S,l}\right)}
{\sum_l \left(3\tau_{T,l}+\tau_{S,l}\right)},
\label{alpha}
\end{equation}
where we added a subscript $l$ to account for different subbands. 
Note that the saturation condition depends
critically on the microwave amplitude $B_{\perp}$, donor spin-flip
rate $\frac{1}{T_1}$ and
resonance linewidth $\frac{1}{T^{*}_{2}}$.  

In transport measurements, SDS is always in competition with other 
scattering sources, such as surface roughness and lattice defects. We 
describe these phenomenologically by adding a rate $\frac{1}{\tau_0}$: 
$\frac{1}{\tau_{S/T}}\rightarrow 
\frac{1}{\tau_0}+\frac{1}{\tau_{S/T}}$.  
Usually, transport is 
dominated by scattering sources other than neutral donors 
($\frac{1}{\tau_{0}}\gg\frac{1}{\tau_{S/T}}$), leading to a reduced 
figure of merit $\alpha \approx \frac{\tau_0}{4} 
\sum_l\left(\frac{1}{\tau_{S,l}}-\frac{1}{\tau_{T,l}}\right)\ll 1$. 
 
The scattering amplitudes to second order in the Born approximation are given 
by 
\begin{eqnarray} 
{\cal A}^{S/T}_{kl,k'l'}&=&\langle \Psi^{S/T}_{kl}| {\cal H}| \Psi^{S/T}_{k'l'}\rangle 
\nonumber\\&&+\sum_{i} \frac{\langle \Psi^{S/T}_{kl}| {\cal H}| i^{S/T}\rangle  
\langle i^{S/T}| {\cal H}| \Psi^{S/T}_{k'l'}\rangle}{(\epsilon_1+\epsilon_{kl})-
(\epsilon_1+\epsilon_i+U)}, 
\label{ast} 
\end{eqnarray} 
and the scattering rates due to a single donor impurity are 
\begin{eqnarray}
\frac{1}{\tau_{S/T,l}}&=&\frac{2\pi}{\hbar}\sum_{k',l'}\left|
{\cal A}^{S/T}_{kl,k'l'}\right|^{2}\left(1-\cos{\theta_{kk'}}\right)\delta(\epsilon_{kl}-\epsilon_{k'l'})
\nonumber\\
&=& \frac{2\pi}{\hbar}\rho\sum_{l'}\left\langle|{\cal 
  A}^{S/T}_{kl,k'l'}|^{2}(1-\cos{\theta_{kk'}})\right\rangle,
\label{taust}
\end{eqnarray}
where $\theta_{kk'}$ is the angle between incoming ($\bm{k}$) and
outgoing ($\bm{k}'$) conduction electron wavevectors, with
$|\bm{k}'|=k$ averaged over the Fermi distribution.  In the degenerate
limit we may set $k$ equal to the Fermi wavevector $k_F$, but a more
general averaging procedure will be considered below.  Here
$\rho=\frac{m^{*}A}{2\pi \hbar^{2}}$ is the 2DEG energy density, with
$A$ the 2DEG area. The energy of the conduction electrons is assumed
to be $\epsilon_{kl}=\epsilon_{\rm{sb},l}+\frac{\hbar^2
  k_{\perp}^{2}}{2m^{*}}$, with $\bm{k}_{\perp}=(k_x,k_y)$, $m^{*}$
the effective mass, and $\epsilon_{\rm{sb},l}$ the ground state energy
of the $l$-th subband.  Eqs.~(\ref{ast})~and~(\ref{taust}) do not
depend explicitly on electron Zeeman energy because we assumed that
the $g$-factor of the donor is the same as the $g$-factor of the
electrons [while it is essential to consider different $g$-factors for
EDMR detection, e.g.  in Eq.~(\ref{weqn}), the $g$-factor differences
give only a negligible effect in the calculation of transport
properties, provided the external field $B$ is not too large].

The second order contribution in Eq.~(\ref{ast}) is a sum over virtual
(intermediate) states, and is generally much larger than the 1st order
contribution.  Fig.~\ref{diagram} illustrates the most important
channels for neutral impurity scattering. These depend on the
two-particle Hamiltonian ${\cal H}$ only through the donor
single-particle orbital energies $\epsilon_i$, donor ``on-site''
Coulomb repulsion $U$, and the overlap between conduction and impurity
wavefunctions.  The largest individual contribution is a virtual
transition to $(\epsilon_1 +U)$, which is a negatively ionized donor
level with two electrons in the ground orbital state
[Fig.~\ref{diagram}(a)]. This transition is only allowed for singlets
due to Pauli exclusion.  Since singlet scattering also has the
channels for excited orbital states available, it is always stronger
than triplet scattering (which has access only to the excited states)
and in general $\alpha > 0$.
 
In these neutral donor scattering events, the total spin of conduction
and donor electrons is always conserved, but their respective spin
states may be exchanged.\cite{bastard90,mahan08} In our notation,
$\Gamma_{\sigma}$ is the rate for a donor with spin $\sigma$ to
exchange spin with a conduction electron having spin $-\sigma$;  using
$\sigma=+,-$ to denote the spin state, the rates can be written as
\begin{eqnarray}
\Gamma_\sigma &=& \frac{2\pi}{\hbar}\sum_{kl,k'l'} \left|J_{kl,k'l'}\right|^{2}
\delta(\epsilon_{kl}-\epsilon_{k'l'})\nonumber\\
&&\times f\left(\epsilon_{kl}+\sigma \frac{g\mu B}{2}\right)
\left[1-f\left(\epsilon_{k'l'}-\sigma\frac{g\mu B}{2}\right)\right].
\quad\phantom{.}
\label{gammasigma}
\end{eqnarray}
Here $J_{kl,k'l'}={\cal A}^{T}_{kl,k'l'}-{\cal A}^{S}_{kl,k'l'}$ are exchange 
scattering amplitudes, and 
\begin{equation}
f(\epsilon)=\frac{1}{\textrm{e}^{\frac{\epsilon-\epsilon_F}{k_BT}}+1}
\end{equation}
are Fermi functions. After converting the sums into integrals over $k$ space, 
the donor spin-flip scattering rate 
becomes  
\begin{equation} 
\frac{1}{T_1}=\Gamma_{\uparrow}+\Gamma_{\downarrow}= \frac{2\pi}{\hbar}\rho^{2}
\left\langle|J_{k,k'}|^2\right\rangle g\mu B \coth{\left(\frac{g\mu B}{2k_BT}\right)}.
\label{t1} 
\end{equation} 
As a check, note that the equilibrium donor spin polarization is given
by $p_{i0}=(\Gamma_{\downarrow}-\Gamma_{\uparrow})
/(\Gamma_{\uparrow}+\Gamma_{\downarrow})= \tanh{\left(\frac{g\mu
      B}{2k_BT}\right)}$.  When $k_BT\gg g\mu B$,
$\frac{1}{T_1}$ scales linearly with the temperature, while in the
opposite regime it scales linearly with $B$. 

When other donor spin-flip mechanisms are active, we must add their
rate to Eq.~(\ref{gammasigma}).  Nevertheless, we will see below that
exchange scattering is the dominant contribution for donors in silicon
in a wide parameter range.

\section{Model calculations and comparison to experiment\label{numerics}}
 
\subsection{Model Hamiltonian and virtual two-particle donor states}

The two-particle Hamiltonian ${\cal H}$ that models the coupling
between a conduction plus a donor impurity electron can be written
explicitly as
\begin{equation}
{\cal H}= {\cal H}_{0}(1)+{\cal H}_{0}(2)+ {\cal C}_{R}(1)+ 
{\cal C}_{R}(2) + {\cal C}_{ee}(1,2)-2\epsilon_F,
\label{htotal}
\end{equation}
where for notational simplicity $(1)$ denotes coordinate $\bm{r}_1$,
and $(2)$ denotes coordinate $\bm{r}_{2}$.  Here ${\cal
  H}_{0}=\hat{T}+V_{z}(z)$ is the translational invariant
single-particle Hamiltonian, with $\hat{T}$ the kinetic energy and
$V_z(z)$ the 2DEG confinement (triangular at low $z$ but flattens out
at high $z$, see Fig.~\ref{band_diagram}). ${\cal C}_R$ is the
attractive Coulomb potential of the impurity, and ${\cal C}_{ee}$ is
the Coulomb electron-electron repulsion,
\begin{subequations}
\begin{eqnarray}
%\hat{T}&=&-\frac{\hbar^{2}}{2m_{t}}\left(\partial^{2}_{x}+\partial^{2}_{y}\right)
% -\frac{\hbar^{2}}{2m_{l}}\partial^{2}_{z},\label{tt}\\
%\hat{V}_z&=&eEz\;\rm{for}\; z>0;\;\hat{V}_z=\infty\; \rm{for}\; z<0, \label{vz}\\
{\cal C}_{R}(i)&=& -\frac{e^{2}}{\kappa}\frac{\textrm{e}^{-q_{\rm{TF}}|\bm{r}_i-\bm{R}|}}
{\left|\bm{r}_i-\bm{R}\right|},\label{cn}\\
{\cal C}_{ee}(1,2)&=&\frac{e^{2}}{\kappa}\frac{\textrm{e}^{-q_{\rm{TF}}|\bm{r}_1-\bm{r}_2|}}
{\left|\bm{r}_{1}-\bm{r}_{2}\right|},\label{cee}
\end{eqnarray}
\end{subequations}
where $\bm{R}=R\hat{z}$ is the location of the donor impurity, and
$\kappa\approx 12$ the dielectric constant for silicon. 
Since we are in the degenerate limit, it is important to account for screening; 
we use Thomas-Fermi screening with wavevector
\begin{equation}
q_{\rm{TF}}=\sqrt{\frac{6\pi e^{2}n_{\rm{2D}}|\phi(z)|^{2}}
{\epsilon_F-\frac{1}{2}(\epsilon_{\rm{sb,1}}+\epsilon_{\rm{sb,2}})}},
\end{equation}
where $n_{\rm{2D}}$ is the 2DEG density and $\phi(z)$ the subband
wavefunction defined below. Note that the last term in
Eq.~(\ref{htotal}) is the chemical potential times the number of
particles.

In order to compute the scattering amplitudes [Eq.~(\ref{ast})], we need
to choose a set of virtual states $|i^{S/T} \rangle$ that forms a
complete basis for the two-particle Hilbert space.  An important
insight is that the conduction electron may hop into the impurity
site, form a negatively ionized virtual donor state, and then hop back
into the Fermi sea.  This motivates the choice of a molecular orbital
basis of negatively ionized donor states:
\begin{equation}
| i^{S/T} \rangle = \frac{1}{\sqrt{2}} 
\left[ | \Phi_1 \Phi_i\rangle \pm | \Phi_i \Phi_1\rangle\right], \;i=1,2,3,\ldots,
\label{twoparti}
\end{equation}
where $\Phi_i(\bm{r})$ is a donor orbital state with single particle
energy $\epsilon_i$, satisfying $\left({\cal H}_{0}+{\cal
    C}_{R}\right)|\Phi_i\rangle = \epsilon_i |\Phi_i\rangle$. Note
that $i=1$ refers to the ground orbital state, and $i=2,3,\ldots$
refers to other excited donor single-electron orbitals.  From
Eq.~(\ref{twoparti}) we see that the state $|1\rangle$ only exists as
a singlet, $|1^{T}\rangle=0$.

The simplest molecular orbital approximation is to assume
\begin{equation}
{\cal H}|i^{S/T}\rangle\approx
(\epsilon_1+\epsilon_i+U-2\epsilon_F)|i^{S/T}\rangle.  
\label{molorb}
\end{equation}
Using this relation, we see that the 2nd order contribution
in Eq.~(\ref{ast}) depends only on the energies $\epsilon_i$, $U$,
and on the overlap integral $S_{kl,i}=\langle \Psi_{kl}|\Phi_i\rangle$. Note
also that states of the type $|\Phi_i\Phi_j\rangle\pm
|\Phi_j\Phi_i\rangle$ with $i,j\neq 1$ do not contribute to
Eq.~(\ref{ast}) [they are orthogonal to Eq.~(\ref{st})]. 

%We remark that each state $|i\rangle^{S/T}$ may be written as a wavepacket of conduction electron states,
%\begin{equation}
%|i\rangle^{S/T}=\sum_{k,k'} \eta^{i}_{k,k'} \left[ |\Psi_{k}\Psi_{k'}\rangle \pm |\Psi_{k'}\Psi_{k}\rangle\right],
%\end{equation}
%because the energy $(\epsilon_1+\epsilon_i+U)$ lie within the band
%continuum (the sum includes $k,k'\gg k_F$). This shows that our basis
%choice is equivalent to an infinite subset of ``positively ionized
%donor levels''. Nevertheless, the choice of a finite basis of
%negatively ionized states is conceptually important to make the
%spin-dependent scattering mechanism physically transparent.

\subsection{Microscopic model for single particle states and overlap integral}

Bulk silicon has a six-fold degenerate conduction band, with energy
minima located at $k_0=0.85\times 2\pi/(5.43~\rm{\AA})$ along the set
of $\langle 100\rangle$ directions in the Brillouin zone. Its
effective mass is anisotropic. Each valley has a heavy mass equal to
$0.98 m_e$ along the valley direction, and a much lighter mass $m^{*}
=0.19 m_e$ in the perpendicular plane ($m_e$ is the free electron
mass).  In the presence of a (001) interface, electrons in the valleys
along $\pm (001)$ will have considerably lower energy, a consequence
of the heavy longitudinal mass.\cite{ando82} This leads to the
following two-subband model for the conduction electron wavefunctions:
$\Psi_{kl}(\bm{r})=\psi_l (z)\textrm{e}^{i\bm{k}\cdot
  \bm{r}_{\perp}}/\sqrt{A}$, with $A$ the 2DEG area and $l=1,2$
a subband label. 
The subband wavefunctions are given by
$\psi_{1}(z)=\phi(z)\sqrt{2}\cos{(k_0 z)}$ and
$\psi_{2}(z)=\phi(z)\sqrt{2}\sin{(k_0 z)}$. 
For $z>0$, we use the Takada-Uemura envelope function, 
\begin{equation}
\phi(z)=\sqrt{\frac{3}{2 z_0}}\left(\frac{z}{z_0}\right)\textrm{e}^{-\frac{1}{2}\left(\frac{z}{z_0}\right)^{3/2}},
\label{env}
\end{equation}
that is known to be an excellent analytic approximation to the
self-consistent 2DEG ground state (see p. 468 of
Ref.~[\onlinecite{ando82}]). We assume $\phi(z)=0$ for $z\leq 0$.  The
characteristic length scale $z_0$ models the 2DEG width, that can be
controlled electrically by the gate voltage.  Our choice of $\psi_l$
corresponds to pure imaginary intervalley scattering at the interface
(other choices of intervalley scattering phase simply give a phase
shift to the cosine and sine functions, that is equivalent to changing
the position of the interface).\cite{sham79} The conduction electron
states are assumed to satisfy ${\cal
  H}_0|\Psi_{k,l}\rangle=\epsilon_{kl}|\Psi_{k,l}\rangle$, where
$\epsilon_{kl}=\frac{\hbar^2}{2m^{*}}k_{\perp}^{2}+\epsilon_{\rm{sb},l}$.
Each subband has energy $\epsilon_{\rm{sb},l}$ at $k=0$, with
valley-splitting
$(\epsilon_{\rm{sb},2}-\epsilon_{\rm{sb},1})/k_B=0-100$~K depending on
interface quality and device geometry.\cite{takashina06,goswami07}

We now discuss model wavefunctions for the donor impurity.  We expect
that donors located too close to the interface will have quite high
$\frac{1}{T_1}$, and the EDMR saturation condition $(\gamma_e
B_{\perp})^2\gtrsim 1/(T_1 T_{2}^{*})$ will not be
satisfied.  For this reason, we expect that 
only donors located far enough from the interface (at depths $R\gg z_0$) can
contribute to EDMR. Therefore it is sufficient to use donor wave
functions that are good approximations in the bulk. Those are the 
Kohn-Luttinger wavefunctions,\cite{kohn57}
\begin{equation}
\Phi_i(\bm{r-R})= F_i (\bm{r-R})\sum_{j=1}^{6} \eta_{ij}
\textrm{e}^{i\bm{k}_j\cdot (\bm{r-R})}.
\label{kohnlutt}
\end{equation}
For simplicity, we used an isotropic envelope $F_i(\bm{r-R})\approx
\textrm{e}^{-r'/a^{*}_i}/\sqrt{\pi a^{*3}_{i}}$, with
$r'^{2}=x^2+y^2+(z-R)^{2}$. The Bohr radius depends on the hydrogenic
principal quantum number $n$ according to $a^{*}_{i}\approx n^2
a^{*}_{1}$ (we used $a^{*}_{1}=18.6$~\AA). The principal quantum
number $n$ relates to the subscript $i$ according to
$n=(i~\rm{div}~6)+1$. Below we will list the donor energy levels with
the zero of energy chosen at flat band (the bulk conduction band
edge).

The valley vectors $\bm{k}_{j}$ are given by $\bm{k}_{1}=+k_0\hat{x}$,
$\bm{k}_{2}=-k_0\hat{x}$, $\bm{k}_{3}=+k_0\hat{y}$,
$\bm{k}_{4}=-k_0\hat{y}$, $\bm{k}_{5}=+k_0\hat{z}$, and
$\bm{k}_{6}=-k_0\hat{z}$. The corresponding $\eta_{ij}$ are conveniently
written as 6-dimensional vectors. Each hydrogenic envelope has six
different valley-split states, classified by symmetry.  The lowest
energy state for Sb has $\epsilon_1= -45$~meV with A1 symmetry,
$\bm{\eta}_{A1}=\frac{1}{\sqrt{6}}(1,1,1,1,1,1)$.\cite{grimeiss82}
This is followed by a three-fold degenerate T1 symmetry level, with
energy $\epsilon_2=-33$~meV and
$\bm{\eta}_{T1}=\frac{1}{\sqrt{2}}(1,-1,0,0,0,0)$,
$\frac{1}{\sqrt{2}}(0,0,1,-1,0,0)$,
$\frac{1}{\sqrt{2}}(0,0,0,0,1,-1)$. Note that only the last state
couples to the (001) subbands. The highest energy level has E symmetry
with energy $\epsilon_3=-31$~meV and is two-fold degenerate,
$\bm{\eta}_{E}=\frac{1}{2}(1,1,-1,-1,0,0)$, or
$\frac{1}{2}(1,1,0,0,-1,-1)$. Again, only one of these couples to
electrons in the lowest energy subband. For principal quantum number
$n\geq 2$ we assumed $\epsilon_n=\epsilon_1/n^{2}$ with
degenerecence $6n^2$. Finally, we used $U=43$~meV for the on-site
Coulomb repulsion.  This corresponds to a binding energy of
$E_{D-}=-(\epsilon_1+U)=2$~meV; this is quite similar to the binding
energy of $1.6$~meV measured for phosphorous (P) in silicon,\cite{narita85}
and calculated by Oliveira and Falicov.\cite{oliveira86} We are not
aware of measurements of $E_{D-}$ for Sb impurities.
 
One important point is that the overlap integral between conduction electrons
in the $l$-th valley and the impurity 
electron $S_{kl,i}=\langle \Psi_{kl}| \Phi_i\rangle$ is strongly 
oscillatory on R. For notational convenience, define $s_1= \cos{(k_0 R)}$, and
$s_2=\sin{(k_0 R)}$, and set
$s_3\equiv s_1$. After a simple calculation we see that 
$S_{kl,i}\propto s_{l}$ apart from a smooth envelope whenever
the valley symmetry of donor orbital $i$ is ``even''
($\propto\cos{[k_0(z-R)]}$), and $S_{kl,i}\propto s_{l+1}$ when the
valley symmetry of orbital $i$ is ``odd'' ($\propto\sin{[k_0(z-R)]}$).
These oscillations occur due to valley interference, 
in a similar fashion as the exchange oscillations between two donor 
impurities in silicon.\cite{cullis70,koiller01,wellard05}

\subsection{Contributions to the scattering amplitude}

We start by evaluating explicitly the first order contributions to
Eq.~(\ref{ast}). To get an idea of the order of magnitude of each
contribution, we will quote numerical values for a single donor
located at $R=6 z_0$, with parameters $A=(1\mu \rm{m})^{2}$,
$\epsilon_{F}=-9.32$~meV,
$\epsilon_{\rm{sb},1}=\epsilon_{\rm{sb},2}=-10$~meV, and $z_0=40$~\AA.
In this case the overlap integral between conduction electron at
$\epsilon_F$ and donor ground state was $S_{k,1}s_1$, with
$S_{k,1}=1.31\times 10^{-4}$.

Let's consider the first order contribution to Eq.~(\ref{ast}). 
Using ${\cal H}_{0}|\Psi_{kl}\rangle = \epsilon_{kl} |\Psi_{kl}\rangle$ and $[{\cal
H}_{0}+{\cal C}_{R}]|\Phi\rangle = \epsilon_1 |\Phi\rangle$, we
get 
\begin{eqnarray}
\langle \Psi^{S/T}_{kl} | {\cal H}|\Psi^{S/T}_{k'l'}\rangle &=& 
\langle \Psi_{kl}| {\cal C}_{R}|\Psi_{k'l'}\rangle +\langle \Psi_{kl}\Phi| {\cal C}_{ee}|\Psi_{k'l'}\Phi\rangle
\nonumber\\
&\pm& \left[
(\epsilon_1 -\epsilon_{F})
S_{k,1}S^{*}_{k',1}s_ls_{l'}\right.\nonumber\\
&& + \langle \Phi |{\cal C}_{R}| \Psi_{k'l'}\rangle
S_{k,1}s_{l'}\nonumber\\&& \left.
+ \langle \Phi \Psi_{kl}|{\cal C}_{12} |\Psi_{k'l'}\Phi\rangle
\right].
\end{eqnarray}
The first contribution on the right hand side is attractive Coulomb
potential scattering (always negative), and the second contribution is
repulsive scattering from the donor electron cloud. This has the same
order of magnitude as the first term, but the opposite sign.  Assuming
$s_l=s_{l'}=1$ (maximum value) these terms are respectively
$-8.1$~neV, and $+8.5$~neV.  The first term inside the brackets equals
$-0.61$~neV, the second $-0.3$~neV, and the third $+0.2$~neV.  Hence
we have $\langle \Psi^{S}_{kl} | {\cal
  H}|\Psi^{S}_{k'l'}\rangle=-0.3$~neV, and $\langle \Psi^{T}_{kl} |
{\cal H}|\Psi^{T}_{k'l'}\rangle=+1.1$~neV. \emph{Interestingly, in
  this particular case the first order contribution to Eq.~(\ref{ast})
  favors triplet scattering}. Actually, computations for different
donor positions $R$ shows that the first order contribution changes
sign several times as the donor position $R$ is varied.

We now turn to second order contributions to Eq.~(\ref{ast}). We
remark that these are always negative for singlets and for triplets
because $\epsilon_1<\epsilon_F < \epsilon_1 +U < \epsilon_i+U$ for
$i=2,3,\ldots$.  The largest individual contribution is a virtual
transition to a negatively ionized donor state with two electrons
occupying the donor ground state. This contributes exclusively to
singlet scattering. Using Eq.~(\ref{molorb}) we get
\begin{equation}
\frac{\langle \Psi^{S}_{kl}| {\cal H}| 1^{S}\rangle
\langle 1^{S}| {\cal H}| \Psi^{S}_{k'l'}\rangle}{\epsilon_{kl}-\epsilon_1-U}= 
\frac{-2(2\epsilon_1 +U -2\epsilon_F)^2\left|S_{k,1}\right|^2 s_l s_{l'}
}{\epsilon_1+U-\epsilon_F},\qquad\phantom{.}
\label{large2nd}
\end{equation}
which equals $-3.7$~neV for the particular case considered. This is
more than $3$ times larger than the 1st order contribution mentioned
above. We checked several other parameter regimes, and found
Eq.~(\ref{large2nd}) to be 2-10 times larger than the 1st order
contribution (singlet or triplet).

All the other second order terms contribute equally to singlet and
triplet. The largest of these involve virtual excited states to the
$n=1$ state with T1 or E symmetry. The T1
state contribution is
\begin{equation}
-3\frac{(\epsilon_1+\epsilon_4+U-2\epsilon_F)^2\left|S_{k,1}\right|^2 s_{l+1} s_{l'+1}}
{\epsilon_4+U-\epsilon_F}=-0.95~\rm{neV},
\end{equation}
and the E contribution is
\begin{equation}
-\frac{3}{2}\frac{(\epsilon_1+\epsilon_6+U-2\epsilon_F)^2\left|S_{k,1}\right|^2 s_{l} s_{l'}}
{\epsilon_6+U-\epsilon_F}=-0.47~\rm{neV}.
\end{equation}

The contributions with principal quantum number $n\geq 2$ may be lumped together in a sum:
\begin{eqnarray}
-\sum_{n=2}^{\infty}\Lambda_n &=& -\sum_{n=2}^{\infty} n^2
\frac{(\epsilon_1 +\epsilon_n +U -2 \epsilon_F)^2}
{\epsilon_n+U-\epsilon_F}\left|S_{k,n}\right|^2 \nonumber\\
&&\times\left(\frac{5}{2}s_l s_{l'}+3s_{l+1}s_{l'+1}\right)
\label{nmax}\\
&\approx& -1.3 \times 10^4 \;\rm{neV},\nonumber
\end{eqnarray}
since there are $3n^2$ orbitals coupling to the 2DEG valleys, each 
with energy $\epsilon_n =
\epsilon_1/n^2$. The envelope of the overlap integrals is defined as 
\begin{equation}
S_{k,n} = \frac{1}{\sqrt{3}}\int d^{3}r \phi(z) 
\frac{\textrm{e}^{i\bm{k}\cdot \bm{r}_{\perp}}}{\sqrt{A}}F_{n}(\bm{r}-\bm{R}).
\end{equation}
We get $\Lambda_2 = 390$~neV, $\Lambda_3 =
6170$~neV, $\Lambda_4 = 4164$~neV, $\Lambda_5 = 1367$~neV, $\Lambda_6
= 425$~neV, $\Lambda_8 = 50$~neV. The $\Lambda_n$ reach a maximum for
$n=3$, and then decrease appreciably with increasing $n$ because the
$S_{k,n}$'s become exponentially small.  As a result, we get a good
approximation by evaluating the sum up to $n_{\rm{max}}=4$. We found
that the contributions for $n\geq 5$ produce negligible changes to our
final result.

Hence the sum of the second order contributions in Eq.~(\ref{ast}) can
be as much as $10^4$ times larger than the first order contributions.
It is interesting to note that the difference between singlet and
triplet rates is determined by the \emph{difference between squared
  amplitudes}: $\frac{1}{\tau_S}-\frac{1}{\tau_T}\propto ({\cal
  A}^{S}_{kl,k'l'})^{2}-({\cal A}^{T}_{kl,k'l'})^{2}$, which is of the
order of $10^4$~neV$^{2}$.  This is several orders of magnitude larger
than the square of the exchange scattering amplitude, that determines
exchange scattering: $\frac{1}{T_1}\propto({\cal
  A}^{T}_{kl,k'l'}-{\cal A}^{S}_{kl,k'l'})^{2}$ is only about $\sim
10$~neV$^{2}$. Therefore, we see that exchange scattering is quite
different than SDS, in the sense that the latter benefits from a large
number of additional electronic transitions.

\subsection{Dependence of EDMR parameters on donor depth}

\begin{figure} 
\includegraphics[angle=-90,width=3in]{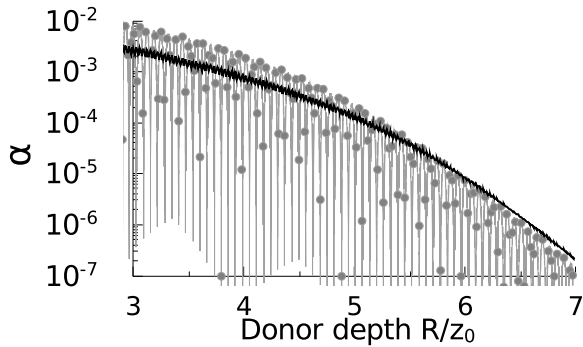} 
\caption{Figure of merit $\alpha$ as a function of donor depth $R$
  divided by the 2DEG thickness $z_0$. Dark curve: Valley degenerate
  case. Grey curve: Valley split case, with
  $\epsilon_{\rm{sb},2}-\epsilon_{\rm{sb},1}\gg k_B T$. Filled circles
  denote actual sites of the diamond lattice, which may be occupied by
  donors.\label{alphafig}}
\end{figure}

\begin{figure} 
\includegraphics[angle=-90,width=3in]{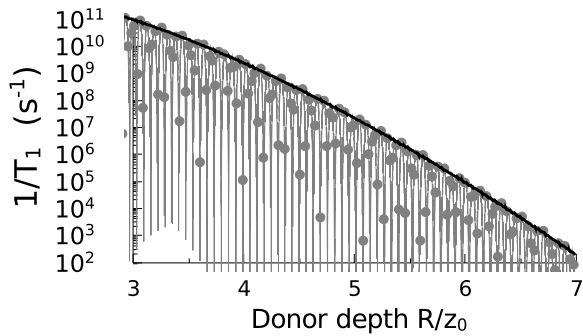} 
\caption{Donor impurity exchange scattering rate $\frac{1}{T_1}$ as a
  function of donor depth $R/z_0$ (same parameters as in
  Fig.~\ref{alphafig}).\label{t1rate}}
\end{figure}

We now show explicit numerical calculations of the parameters
determining EDMR detection, and discuss their dependence with donor
depth $R$. The figure of merit $\alpha$ is shown in
Fig.~\ref{alphafig}, and the donor spin flip rate $\frac{1}{T_1}$ due
to the exchange scattering mechanism is shown in Fig.~\ref{t1rate}.
Both $\alpha$ and $\frac{1}{T_1}$ decrease appreciably as $R$
increases relative to the 2DEG thickness $z_0$, and are quite
sensitive to $\epsilon_F-\epsilon_{\rm{sb},l}$ (or, equivalently, the
2DEG area density). We used
$\epsilon_F-\epsilon_{\rm{sb,1}}=0.68$~meV, and considered two
different cases: Valley degenerate with
$\epsilon_{\rm{sb,1}}=\epsilon_{\rm{sb},2}=-25$~meV, and valley-split
with $\epsilon_{\rm{sb},2}=-15$~meV and
$\epsilon_{\rm{sb},1}=-25$~meV.  Both $\alpha$ and $\frac{1}{T_1}$ are
independent of the 2DEG area $A$; $\alpha$ depends sensitively on the
scattering time due to other sources; we used $\tau_0=0.4$~ps which is
equivalent to a typical transistor mobility of $4\times
10^{3}$~cm$^{2}/(\rm{Vs})$ ($\frac{1}{T_1}$ is independent of
$\tau_0$). We assumed $T=5$~K and $B=0.36$~T ($\alpha$ does not depend
on $B$, and is nearly independent of $T$ in the valley degenerate
case).

Fig.~\ref{t1rate} shows that exchange scattering is several orders of
magnitude stronger than conventional spin-phonon coupling of isolated
donor impurities. For example, Feher and Gere\cite{feher59} measured
$\frac{1}{T_1}\sim 10^{-1}$~s$^{-1}$ for phosphorous donors in bulk
silicon at $T=5$~K and $B=0.36$~T.

Interestingly, the EDMR amplitude depends on whether the device
temperature is larger than the valley-splitting energy or not.  At
large temperatures (or small valley splittings), $\alpha$ and
$\frac{1}{T_1}$ are smooth functions of $R$ (apart from tiny 
oscillations), because contributions from the two subbands complement
each other (dark curves). For the opposite regime of temperature lower
than valley-splitting, the parameters are strongly oscillatory on $R$
(grey curves). In this case $\frac{1}{T_1}$ becomes quite small
for some donor positions close to the interface, suggesting that a
fraction of the donors located at ``lucky sites'' might be detectable
by EDMR. In Figs.~\ref{alphafig}~and~\ref{t1rate} we used filled
circles to denote the actual silicon lattice sites that may be
occupied by substitutional donors.

\begin{table} 
\begin{center} 
\begin{tabular}{c |c |c | c } 
 &    & Experiment & Theory \\ 
$n_{\rm{2DEG}}$ ($10^{11} \rm{cm}^{-2}$) &  $B_{\perp}$ (G) &  $\frac{|\Delta 
I|}{I_0}$ ($10^{-8}$)  &  $\frac{|\Delta I|}{I_0}$ ($10^{-8}$) \\ 
\hline 
1.1  &  0.3  & 10 & 5.6 \\ 
%\hline 
2.2 &  0.3  & 4.0 & 2.2 \\ 
%\hline 
3.8  & 0.3   & 2.0  & 0.5 \\ 
\hline 
2.2 &  0.07  & 1.0 & 0.5 \\ 
%\hline 
2.2 &  0.14  & 2.5 & 1.0 \\ 
%\hline 
2.2 & 0.19  & 3.5  & 1.4 \\ 
%\hline 
2.2 & 0.28   & 4.0  & 2.1 \\ 
%\hline 
2.2 & 0.55   & 5.0  & 3.8 \\ 
\hline 
\end{tabular} 
\caption{Comparison between theory and experiment\cite{lo07} at 
  $T=5$~K and $B=0.36$~T with an inhomogeneously broadened linewidth 
  of $\frac{1}{\gamma_e T_{2}^{*}}=2$~G.  Our theoretical calculations of 
$\Delta I/I_0$ were obtained by multiplying the donor density by the 
scattering rate due to a single donor, and integrating over donor depth.  We 
used 2DEG thickness $z_0=50$~\AA, valley-degenerate 
  subband energy $\epsilon_{\rm{sb},1}=\epsilon_{\rm{sb},2}=-25$~meV (zero energy is at flat 
  band), and transistor mobility $4\times 10^3$~cm$^2$/Vs (corresponds 
  to $\tau_0=0.4$~ps).} 
\label{table1} 
\end{center} 
\end{table} 

\subsection{Comparison to experiment} 
 
Our theory can be directly compared to experiments. Ghosh and
Silsbee\cite{ghosh92} measured $\Delta
I/I_0=(I_{\rm{off~res.}}-I_{\rm{at~res.}})/I_0 \sim - 10^{-5}$ at high
power (the current increased upon resonance), while at lower power
they found that $\Delta I$ changed sign.  In this study, a silicon
transistor bulk doped with phosphorous impurities was used.  Lo {\it
  et al.}\cite{lo07} measured much lower amplitudes $|\Delta I|/I_0=
10^{-8}-10^{-7}$, with antimony donors \emph{implanted only into the
transistor channel}. Unfortunately the use of derivative detection did
not allow the determination of the sign of $\Delta I$.
 
Our theory disagrees in sign with the measurements of
[\onlinecite{ghosh92}], suggesting that their high power signal was
not due to the SDS mechanism, instead it was likely due to the 2DEG
heating mechanism.\cite{stein83}

We now compare our explicit numerical results with the experimental
data of Lo {\it et al}.\cite{lo07} Our theory allows the prediction of
the device current for a given donor distribution \emph{without any
  fitting parameters}.  Denote $N_j=A\Delta z \;n_d(z_j)$ the number
of donor impurities located in the interval $\Delta z=z_{j+1}-z_{j}$,
with $n_d(z_j)$ the volume density of donors in the $\Delta z$-thick
layer. The EDMR signal averaged over the donor profile
is simply given by
\begin{equation}
\frac{I-I_0}{I_0} = p_c\sum_j N_j\alpha(z_j)p_i(z_j).
\label{iavg}
\end{equation}
We obtained the depth dependent donor density $n_d(z)$ from Secondary
Ion Mass Spectroscopy measurements (SIMS).  Our explicit model
calculations of the EDMR current are shown in Table~\ref{table1} for
all measurements made in [\onlinecite{lo07}].  The EDMR response
predicted by theory is in reasonable agreement with experiment.

% Begin discussion of optimization 

\subsection{Physical optimization of EDMR}
 
\begin{figure} 
\includegraphics[angle=-90,width=3in]{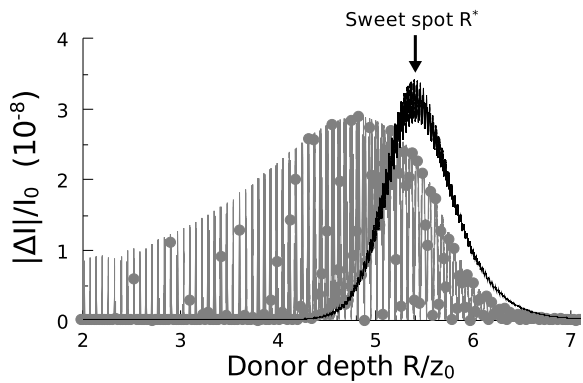} 
\caption{EDMR amplitude at $T=5$~K as a function of donor placement, 
  for zero valley-splitting (dark curve) and for 
  valley-splitting equal to $100$~K (oscillating gray curve, with 
  filled circles denoting actual donor sites of the diamond 
  lattice).\label{current_z}} 
\end{figure} 
 
Table~\ref{table1} shows that the SDS contribution to EDMR is much
weaker than anticipated on the basis of previous
measurements.\cite{ghosh92} We now describe a physical optimization
that aims at maximizing $|\Delta I|/I_0$ for the minimal number of
donor impurities.
 
The SDS parameters $\alpha$ and $\frac{1}{T_1}$ decrease rapidly with 
increasing donor depth $R$, an effect that has important implications 
for the optimization of the EDMR amplitude. Notably, for given  
$B_{\perp}$, the EDMR signal will be maximal for donors satisfying 
$\frac{1}{T_1}(R^*)\approx T_{2}^{*}(\gamma_e B_{\perp})^{2}$.  The location 
$R^*$ is the closest one to the interface that satisfies  
the saturation condition $(\gamma_e B_{\perp})^{2}\gtrsim 
1/(T_1T^{*}_{2})$. 
 
Fig.~\ref{current_z} shows the EDMR amplitude as a function of donor 
location, for a \emph{single donor implanted in the center of a transistor of 
area 
$0.1(\mu\rm{m})^{2}$}.  We assumed microwave amplitude $B_{\perp}= 
0.3$~G, 2DEG density $1.1\times 10^{11}$~cm$^{2}$, with other 
parameters as in Table I. 
 
Interestingly, the EDMR amplitude depends on whether the device 
temperature is larger than the valley-splitting energy or not.  At 
large temperatures, $|\Delta I|/I_0$ is a smooth function of $R$, 
because contributions from the two subbands complement each other. In 
this case a single donor placed at the sweet spot $R^*\approx 5.5~z_0$ 
is optimally detected by EDMR (dark curve in Fig.~\ref{current_z}, with 
$\epsilon_{\rm{sb1}}=\epsilon_{\rm{sb2}}=-25$~meV).  For the opposite 
regime of temperature lower than valley-splitting, $|\Delta I|/I_0$ is 
strongly oscillatory on $R$, and donors closer to the interface 
can be detected as well (oscillating gray curve in Fig.~\ref{current_z}, with 
$\epsilon_{\rm{sb1}}=-25$~meV and $\epsilon_{\rm{sb2}}=-15$~meV). 
 
Fig.~\ref{current_z} shows that EDMR is able to detect a single donor spin 
implanted in a transistor of area $0.1 (\mu\rm{m})^2$. For 
typical $I_0\sim 5$~$\mu$A, current modulations $\Delta I\sim 0.1$~pA 
are detectable with standard techniques, \emph{provided the donor is 
  placed at the sweet spot $R^*$}. A single donor resonance may be 
identified by the presence of only one hyperfine satellite line out of 
the full satellite spectrum with 2I+1 lines (for donor nuclear spin 
I), for measurements within the nuclear spin-flip time.\cite{sarovar07}
 
The EDMR amplitude is directly proportional to the donor area density 
per monolayer.  Therefore, if $6\times 10^6$ Sb donors were to be 
placed exactly at the sweet spot $R^*$ in a large transistor of area 
$\sim 10^3 (\mu\rm{m})^{2}$, the signal would be $|\Delta I|/I_0\sim 
10^{-5}$, two orders of magnitude higher than in [\onlinecite{lo07}]. 

\section{Kondo temperature and validity of perturbation theory\label{kondosection}}

Our approach is based on perturbation theory
[Eq.~(\ref{ast})]. This is known to be a good approximation only when the temperature
is larger than the characteristic Kondo temperature $T_{\rm{Kondo}}$.
For our problem, the Kondo temperature may be written as\cite{haldane78}
\begin{equation}
k_B T_{\rm{Kondo}}=\sqrt{\left(\epsilon_1+U-\epsilon_F\right)
\left(\epsilon_F-\epsilon_{\rm{sb}}\right)\left|J\rho\right|}\textrm{e}^{-\frac{1}{|J\rho|}},
\label{kondo}
\end{equation}
where $J=\langle J_{kl,k'l'}\rangle$ is an average exchange
energy at the Fermi level (averaged over subbands), and
$\epsilon_{\rm{sb}}$ is taken as the lowest subband energy.
Eq.~(\ref{kondo}) applies when $\epsilon_1+U>\epsilon_F$ and
$\epsilon_1<\epsilon_{\rm{sb}}<\epsilon_F$.  The first term in the
square root is a characteristic particle excitation bandwidth for
electrons tunneling into the donor, while the second term is a
hole excitation bandwidth. 

\begin{figure} 
\includegraphics[angle=-90,width=3in]{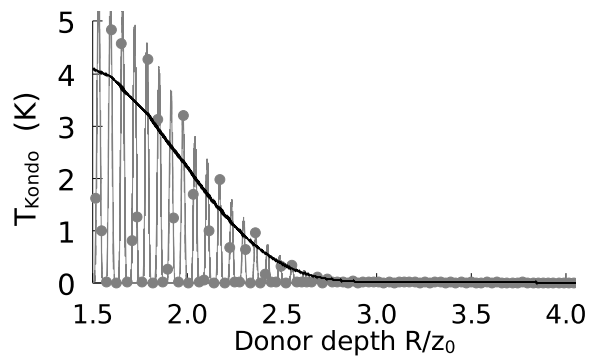} 
\caption{Kondo temperature as a function of donor depth $R/z_0$ (same
  parameters as in Fig.~\ref{alphafig}). Note that the Kondo temperature
  is exponentially small for $R>3 z_0$; this justifies our
  perturbative approach.\label{kondofig}}
\end{figure}

Fig.~\ref{kondofig} shows our calculated Kondo temperature as a
function of donor depth. Here we see that for $R>3 z_0$, we have
$T_{\rm{Kondo}}\ll 0.1$~K. Hence, even at the lowest temperatures
achievable experimentally, there is no Kondo effect for donors located
at $R>3 z_0$ ($T\gg T_{\rm{Kondo}}$).

In the valley degenerate regime, only donors around $R\sim 5.5 z_0$
can be detected by EDMR, as shown in Fig.~\ref{current_z} (donors
closer to the interface can be detected at higher power).  Hence we
see that donors located in this region satisfy $T\gg T_{\rm{Kondo}}$
for the lowest temperatures achievable in the laboratory, and our
perturbation theory approach is justified. In the valley degenerate
case, both the EDMR current as well as the Kondo temperature are
strongly oscillatory with donor depth, and again it can be seen that
all donors that are detectable by EDMR have extremely low
$T_{\rm{Kondo}}$. Note that the behavior of $T_{\rm{Kondo}}$ as a
function of donor depth is qualitatively similar to exchange
scattering; as a consequence, all donors with $\frac{1}{T_1}$ low
enough to be detectable by EDMR have a corresponding quite small
$T_{\rm{Kondo}}$.

\section{Spin injection \label{spininjec}}

We now discuss the spin injection regime. Under spin injection, each
spin species have different quasi-Fermi energies, $\epsilon_{F\uparrow}\neq
\epsilon_{F\downarrow}$.\cite{zutic04} 
For $k_B T\ll
\rm{Min}\{\epsilon_{F\sigma}-\epsilon_{\rm{sb,l}}\}$, the carrier
polarization is well approximated by 
\begin{equation}
p_c =\frac{n_{\uparrow}-n_{\downarrow}}{n_{\uparrow}+n_{\downarrow}}
\approx\frac{2\left(\epsilon_{F\uparrow}-\epsilon_{F\downarrow}+g\mu
B\right)}{\sum_l \left(\epsilon_{F\uparrow}
+\epsilon_{F\downarrow}-2\epsilon_{\rm{sb,l}}\right)}.
\end{equation}
Here $n_{\sigma}=\int \rho d\epsilon f_{\sigma}(\epsilon)$ is the 2DEG
area density for each spin subband. 

The impurity scattering times
$\tau_{S/T}$ must now be calculated for each spin subband:
\begin{equation}
\left\langle \tau_{S/T,l}\right\rangle_{\sigma} = \frac{\int \rho
  d\epsilon \left(\epsilon-\epsilon_{\rm{sb},l}+\sigma\frac{g\mu
  B}{2}\right)\left(-\frac{\partial f_{\sigma}}{\partial
  \epsilon}\right)\tau_{S/T,kl}}
{\int \rho d\epsilon
  \left(\epsilon-\epsilon_{\rm{sb},l}
+\sigma\frac{g\mu B}{2}\right)\left(-\frac{\partial f_{\sigma}}
{\partial \epsilon}\right)},
\end{equation}
where we added a spin subscript to the scattering times and the Fermi functions.
The energy integrals are from
$(\epsilon_{\rm{sb},l}-\sigma\frac{g\mu B}{2})$ to $\infty$. The current
is now calculated as in Eq.~(\ref{mu}), using $p_S=(1-p_i)/4$ for the
spin up subband, and $p_S=(1+p_i)/4$ for the spin down subband, with
$p_T=1-p_S$ in each case. In addition, each $\langle
\tau_{S/T}\rangle_{\sigma}$ must be multiplied by its 
corresponding density $n_{\sigma}$.
The source-drain current becomes 
\begin{equation}
\frac{I-I_0}{I_0}=\frac{\sum_{l}n_{\uparrow}\left(\langle
    \tau_{Tl}\rangle_\uparrow-\langle\tau_{Sl}\rangle_\uparrow\right)
-n_{\downarrow}\left(\langle
    \tau_{Tl}\rangle_\downarrow-\langle\tau_{Sl}\rangle_\downarrow\right)}
{\sum_{l}n_{\uparrow}\left(3\langle
    \tau_{Tl}\rangle_\uparrow+\langle\tau_{Sl}\rangle_\uparrow\right)
+n_{\downarrow}\left(3\langle
    \tau_{Tl}\rangle_\downarrow+\langle\tau_{Sl}\rangle_\downarrow\right)}p_i.
\label{dioiinj}
\end{equation}
Interestingly, under spin injection the scattering times are
intertwined with the densities $n_{\sigma}$, and a figure of merit
$\alpha$ independent of $p_c$ can not be defined.  Note that when
$\langle \tau_{S/T}\rangle_{\uparrow}=\langle
\tau_{S/T}\rangle_{\downarrow}$, Eq.~(\ref{dioiinj}) becomes $\Delta
I/I_0 = \alpha p_c p_i$, as obtained previously.

The donor spin transition rates are also modified. Following from
Eq.~(\ref{gammasigma}), $\Gamma_\sigma\propto \int d\epsilon
f_{-\sigma}(\epsilon)[1-f_{\sigma}(\epsilon -\sigma \frac{g\mu
  B}{2})]$, leading to
\begin{eqnarray}
\frac{1}{T_1}&=& \frac{2\pi}{\hbar}\rho^{2}
\left\langle|J_{kl,k'l'}|^2\right\rangle  
\left(\epsilon_{F\uparrow}-\epsilon_{F\downarrow}+g\mu B\right)\nonumber\\
&&\times\coth{\left(\frac{\epsilon_{F\uparrow}-\epsilon_{F\downarrow}+g\mu B}{2k_BT}\right)}.
\label{t1inj}
\end{eqnarray}
The steady state impurity
polarization becomes 
\begin{equation}
p_{i0}=\tanh{\left(\frac{\epsilon_{F\uparrow}-\epsilon_{F\downarrow}+g\mu
      B}{2k_B T}\right)}.
\label{pi0inj}
\end{equation}

\begin{figure} 
\includegraphics[angle=-90,width=3in]{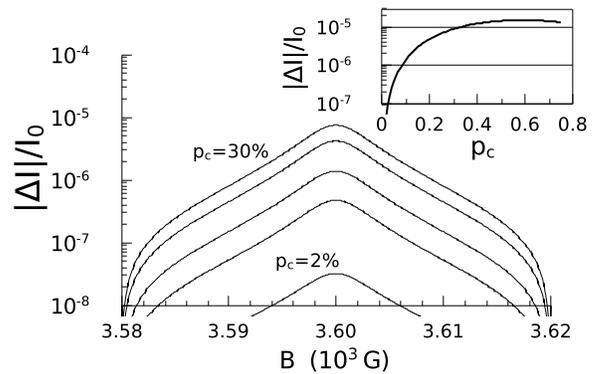} 
\caption{EDMR lineshape under spin injection, for carrier spin polarization 
  $p_c=2,6,10,20,30$\%.  Inset: EDMR amplitude at resonance, as a 
  function of $p_c$.\label{current_pc}} 
\end{figure}

Fig.~\ref{current_pc} shows EDMR lineshapes under spin injection,
using $T=5$~K, $B_{\perp}=0.3$~G, and doping profile and other
parameters similar to [\onlinecite{lo07}].  The 2DEG density is fixed
at $2\times 10^{11}$~cm$^{-2}$, and the conduction electron spin
polarization is varied between 2\% and 75\%. At low $p_c$, the EDMR
amplitude scales as $p_c p_{i0}\propto p_{c}^{2}$, since $\alpha$
remains unchanged and $p_{i0}$ scales proportional to $p_c$ due to
exchange scattering [Eq.~(\ref{pi0inj})].  In this regime, EDMR can be
used as a local probe of carrier spin polarization.
 
However, as $p_c$ is increased beyond the threshold 
\begin{equation}
p_T = \frac{\textrm{Max} \{2k_BT,g\mu B\}}
{(\epsilon_{F\uparrow}+\epsilon_{F\downarrow})-(\epsilon_{\rm{sb1}}+\epsilon_{\rm{sb2}})},
\end{equation}
the spin-flip rate starts increasing as $\frac{1}{T_1}\propto p_c$,
because any donor spin antiparallel to the 2DEG spins will relax
rapidly [note that Eq.~(\ref{t1inj})
becomes proportional to $p_c$ when $p_c>p_T$].  By virtue of these
larger $\frac{1}{T_1}$'s, EDMR will excite donors further away from
the interface, tending to decrease $\Delta I/I_0$.  Remarkably,
this effect competes against the $p_{c}^{2}$ scaling, saturating
$\Delta I/I_0$ for $p_c\gg p_T$. While this limits EDMR as a probe of
carrier spin for $p_c\lesssim p_T$, it also demonstrates that EDMR
detection is optimal at $p_c\approx p_T$.

\section{Conclusions \label{conclusions}}

In conclusion, we presented a microscopic theory of spin-dependent
scattering in the interaction of conduction electrons with neutral
donor atoms. Our results are based on an effective mass approximation.
More sophisticated approaches are likely to reduce the valley
oscillations,\cite{wellard05} with no modification to our conclusions.
The considered mechanism requires temperatures lower than the impurity
binding energy ($T<100$~K), but even higher temperatures may be
achieved using deep level magnetic atoms or clusters. 

We showed that SDS is determined by virtual transitions into doubly
occupied donor states. As a result, SDS always leads to a reduced
current upon EDMR saturation, since $\alpha>0$. 

A recent experiment\cite{beveren08} demonstrates that the EDMR current
due to SDS is indeed reduced upon donor spin saturation (Fig.~2(c) in
[\onlinecite{beveren08}]), in agreement with our proposed virtual
transition mechanism.

Spin-dependent scattering detection is challenging due to competing
heating effects,\cite{ghosh92} which are directly proportional to the
number of donors present.  Our finding that SDS arises solely from
impurity spins located within a narrow depth window with respect to
the 2DEG shows a path to significant optimization of spin signal
intensity for a minimal number of donors placed into this depth
window, and underpins the development of single spin readout devices.

Our theory shows how the EDMR amplitude will scale with carrier spin
polarization in the regime of spin
injection.\cite{zutic06,jonker07,appelbaum07} Therefore, the
monitoring of donor electron spin resonances can be utilized for the
spatially resolved characterization of conduction electron spin
polarization, providing a sensitive probe for optimization of spin
injection and spin transport in semiconductors with indirect band gap
and weak spin-orbit coupling.
 
We thank M. Friesen, T. Schenkel, A.M. Tyryshkin, and I. 
\v{Z}uti\'{c} for a careful reading of the manuscript, and A.L. Efros, 
X. Hu, B. Koiller, S.A.  Lyon, I.  Martin, J.E. Moore, and A.G. 
Petukhov for useful discussions.  RdS acknowledges support from NSERC 
and the UVic Faculty of Sciences; CCL and JB acknowledge support from 
WIN and NSA.


\begin{thebibliography}{99} 
 
\bibitem{zutic04} I. \v{Z}uti\'{c}, J. Fabian, and S. Das Sarma, \rmp 
  {\bf 76}, 323 (2004). 
 
\bibitem{kane98} B.E. Kane, Nature {\bf 393}, 133 (1998). 
 
\bibitem{zutic06} I. \v{Z}uti\'{c}, J. Fabian, and S.C. Erwin, \prl {\bf 97}, 
  026602 (2006). 
   
\bibitem{tyryshkin03} A.M. Tyryshkin, S.A. Lyon, A.V. Astashkin, and 
  A.M. Raitsimring, \prb {\bf 68}, 193207 (2003). 
 
\bibitem{kikkawa99} J.M. Kikkawa and D.D. Awschalom, Nature {\bf 397}, 
  139 (1999).

\bibitem{crooker05} S.A. Crooker, M. Furis, X. Lou, C. Adelmann, D.L.
  Smith, C.J. Palmstr\o m, and P.A. Crowell, Science {\bf 309}, 2191
  (2005).
 
\bibitem{jonker07} B.T. Jonker, G. Kioseoglou, A.T. Hanbicki, C.H. Li 
  and P.E. Thompson, Nature Physics {\bf 3}, 542 (2007). 
   
\bibitem{appelbaum07} I. Appelbaum, B. Huang, and D.J. Monsma, Nature 
  {\bf 447}, 295 (2007). 
 
\bibitem{ghosh92} R. N. Ghosh and R.H. Silsbee, \prb {\bf 46}, 12508 (1992). 
 
\bibitem{lo07} C.C. Lo, J. Bokor, T. Schenkel, J. He, A.M. Tyryshkin, 
  and S.A. Lyon, Appl. Phys. Lett. {\bf 91}, 242106 (2007). 
 
\bibitem{xiao04} M. Xiao, I. Martin, E. Yablonovitch, and H.W. Jiang,
  Nature {\bf 430}, 435 (2004).

\bibitem{elzerman04} J.M. Elzerman, R. Hanson, L.H. Willems van
  Beveren, B. Witkamp, L.M.K. Vandersypen, L.P. Kouwenhoven, Nature
  {\bf 430}, 431 (2004).
 
\bibitem{lepine72} D.J. Lepine, \prb {\bf 6}, 436 (1972).

\bibitem{kaplan78} D. Kaplan, I. Solomon, and N.F. Mott, J. Phys.
  (Paris), Lett. {\bf 39}, L51 (1978).

\bibitem{christmann95} P. Christmann, W. Stadler, and B.K. Meyer, \apl
  {\bf 66}, 1521 (1995).

\bibitem{brandt04} M.S. Brandt, S.T.B. Goennenwein, T. Graf, H. Huebl,
  S. Lauterbach, and M. Stutzmann, phys. stat. sol. (c) {\bf 1}, 2056 (2004).

\bibitem{stegner06} A.R. Stegner, C. Boehme, H. Huebl, M. Stutzmann,
  K. Lips, and M.S. Brandt, Nature Physics {\bf 2}, 835 (2006).

\bibitem{mccamey06} D.R. McCamey, H. Huebl, M.S. Brandt, W.D.
  Hutchison, J.C. McCallum, R.G. Clark, and A.R. Hamilton, \apl {\bf
    89}, 182115 (2006).

\bibitem{morley08} G.W. Morley, D.R. McCamey, H.A. Seipel,
  L.-C. Brunel, J. van Tol, and C. Boehme, \prl {\bf 101}, 207602 (2008).

\bibitem{honig66} A. Honig, \prl {\bf 17}, 186 (1966). 

\bibitem{schmidt66} J. Schmidt and I. Solomon, C.R. Acad. Sci. B {\bf 263}, 169 (1966).
 
\bibitem{sarovar07} M. Sarovar, K.C. Young, T. Schenkel, and K.B. 
  Whaley, \prb {\bf 78}, 245302 (2008).
 
\bibitem{bastard90} G. Bastard and L.L. Chang, \prb {\bf 41}, 7899 (1990).

\bibitem{mahan08} G.D. Mahan and R. Woodworth, \prb {\bf 78}, 075205 (2008).

\bibitem{haldane78} F.D.M. Haldane, J. Phys. C {\bf 11}, 505 (1978).

\bibitem{ando82} T. Ando, A.B. Fowler and A. Stern, \rmp {\bf 54}, 437 
(1982). 
   
\bibitem{sham79} L.J. Sham and M. Nakayama, \prb {\bf 20}, 734 (1979).

\bibitem{takashina06} K. Takashina, Y. Ono, A. Fujiwara, Y. Takahashi, 
  and Y. Hirayama, \prl {\bf 96}, 236801 (2006). 

\bibitem{goswami07} S. Goswami, K.A. Slinker, M. Friesen, L.M.
  McGuire, J.L. Truitt, C. Tahan, L.J. Klein, J.O. Chu, P.M. Mooney,
  D.W. van der Weide, R. Joynt, S.N. Coppersmith and M.A. Eriksson,
  Nature Physics {\bf 3}, 41 (2007).
   
\bibitem{kohn57} W. Kohn, {\it Solid State Physics}, Vol. {\bf 5}, 257 (1957).

\bibitem{grimeiss82} H.G. Grimmeiss, E. Janz\'{e}n, and K. Larsson, 
  \prb {\bf 25}, 2627 (1982).
  
\bibitem{narita85} S. Narita, Solid State Commun. {\bf 53}, 1115
  (1985).

\bibitem{oliveira86} L.E. Oliveira and L.M. Falicov, \prb {\bf 33}, 6990 (1986).
 
\bibitem{cullis70} P.R. Cullis and J.R. Marko, \prb {\bf 1}, 632 (1970).

\bibitem{koiller01} B. Koiller, X. Hu and S. Das Sarma, \prl {\bf 88}, 
  027903 (2001).

\bibitem{wellard05} C.J. Wellard and L.C.L. Hollenberg, \prb {\bf 72}, 085202 (2005).
   
\bibitem{feher59} G. Feher and E.A. Gere, Phys. Rev. {\bf 114}, 1245
  (1959). 

\bibitem{stein83} D. Stein, K.v. Klitzing, and G. Weimann, \prl {\bf 51}, 130 
(1983). 

\bibitem{beveren08} L.H. Willems van Beveren, H. Huebl, D.R. McCamey, T.
  Duty, A.J. Ferguson, R.G. Clark, and M.S. Brandt, \apl {\bf 93},
  072102 (2008).

\end{thebibliography}
\end{document}